# NETWORK STRUCTURE AND COLLECTIVE INTELLIGENCE IN THE DIFFUSION OF INNOVATION

Joshua Becker

UCL School of Management

University College London

joshua.becker@ucl.ac.uk

**ABSTRACT**

When multiple innovations compete for adoption, historical chance leading to early advantage can generate lock-in effects that allow suboptimal innovations to succeed at the expense of superior alternatives. Research on the diffusion of innovafacetion has identified many possible sources of early advantage, but these mechanisms can benefit both optimal and suboptimal innovations. This paper moves beyond chance-as-explanation to identify structural principles that systematically impact the likelihood that the optimal strategy will spread. A formal model of innovation diffusion shows that the network structure of organizational relationships can systematically impact the likelihood that widely adopted innovations will be payoff optimal. Building on prior diffusion research, this paper focuses on the role of central actors i.e. well-connected people or firms. While contagion models of diffusion highlight the benefits of central actors for spreading innovations further and faster, the present analysis reveals a dark side to this influence: the mere presence of central actors in a network increases rates of adoption but also increases the likelihood of suboptimal outcomes. This effect, however, does not represent a speed-optimality tradeoff, as dense networks are both fast and optimal. This finding is consistent with related research showing that network centralization undermines collective intelligence.

# NETWORK STRUCTURE AND COLLECTIVE INTELLIGENCE IN THE DIFFUSION OF INNOVATION

## INTRODUCTION

Explaining how suboptimal innovations can become widely adopted at the expense of superior alternatives, as for example with competing technologies, is an important question for research on the diffusion of innovation (Greve & Seidel, 2015). When adopters prefer to make the same choice as peers, the widespread adoption of any option including suboptimal choices can generate lock-in effects that prevent superior alternatives from spreading (Arthur, 1989; Greve & Seidel, 2015). As a result, the diffusion of innovation is path dependent, meaning that chance historic events giving some lucky option an early advantage leads to an essentially random outcome (Arthur, 1989; Greve & Seidel, 2015; Loch & Huberman, 1999; Puffert, 2002; Sydow, Schreyogg, & Koch, 2009). Theoretical and empirical research has revealed many mechanisms such as novelty or market positioning that can provide this initial strategic advantage (Damanpour, 1987; Lee, Smith, & Grimm, 2003; Suarez, Grodal, & Gotsopoulos, 2015). One key finding comes from network analysis, which reveals that central actors can spread innovations further and faster (Banerjee, Chandrasekhar, Duflo, & Jackson, 2013; Banerjee et al., 2013; Coleman, Katz, & Menzel, 1957; Soh, 2010; Watts & Dodds, 2007).

However, the explanation of randomness or even first-mover advantage just punts the problem further down the logical line, and begs the question of why it is sometimes the optimal innovation and sometimes a suboptimal choice that obtains the initial advantage. Because the mechanisms of initial advantage can help optimal and suboptimal innovations alike, this prior research can guide firms seeking to promote some particular technology but offers no insight into what structural features of a market as a whole impact the likelihood that the diffusion process will surface the best innovation. In particular, the observation that central actors can benefit individual innovations does not inform whether the presence or absence of central actors as a structural feature will increase or decrease the optimality of innovation diffusion. On the one hand, central actors bridge disparate parts of the network and thus may be more effective at finding optimal solutions (Clement, Shipilov, & Galunic, 2018) such that their influence improves optimality. On the other hand, central actors can be highly influential (Soh, 2010; Watts & Dodds, 2007) and may thus crowd out potentially superior solutions before they surface.

This paper examines how network structure impacts the likelihood of optimal diffusion when multiple innovations compete for adoption. This paper contributes to research on the diffusion of innovation by showing how path dependent processes are not just random—it is possible to identify structural features of an industry or market that make it more or less likely to converge on optimal innovations. Prior research has shown how computational models can be used to examine the effect of network structure on the diffusion of innovation, but has focused on the strength of lock-in effects and the speed of diffusion when a new innovation is introduced to an established market (Borgatti & Foster, 2003; Davis & Greve, 1997; Ellison, 1993; Lee, Song, & Yang, 2016). In contrast, the present analysis studies game theoretic equilibrium selection in a diffusion model to examine how the presence of central actors in a network impacts the likelihood that the group will select the optimal choice. Results show that centralized networks characterized by the presence of disproportionately central actors are more likely to converge on suboptimal innovations. However, these results do not reflect a speed-optimality tradeoff, and decentralized networks can be both fast and optimal.

**The Diffusion of Innovation and the Emergence of Conventions**

The population-wide adoption of a technology emerging from distributed individual decision-making is an example of a more general social behavior known as a “convention” (Lewis, 1969; Young, 1993). Prior research has shown both game theoretically and empirically how conventions can emerge spontaneously in a population simply due to individual preferences to match peers (Centola & Baronchelli, 2015; Young, 1993). When adoption choice is shaped by matching incentives such that adopters want to make the same choice as their peers, the value of any given option increases as its rate of adoption increases. Once any option becomes widespread, these increasing returns to adoption mean that popularity generates lock-in effects meaning that even superior alternatives will no longer be adopted (Arthur, 1989). This lock-in occurs because conventions are self-reinforcing: nobody wants to use a technology such as a messaging service if it’s not compatible with anybody else’s choice, which would mean that they have nobody to message. Moreover, any other choice would be “unconventional.”

The phenomena of interest in the present paper are innovations where both payoff and matching incentives are a factor, such that conventions are likely to emerge and optimization is a concern—the intersection of theory on the diffusion of innovation and the emergence conventions. This focus includes innovation diffusion where adopters must consider both the value of each choice and the popularity among peers. For example, early railroad companies had a strong interest in choosing railroad gauge that was compatible with geographically proximate companies in order to facilitate the joining of their networks, but also had to consider performance attributes of different choices (Puffert, 2002). Software developers have an interest in using the same programming language as their peers, in order to facilitate resource and knowledge sharing, but different languages offer different features (Fang, Wu, & Clough, 2021).

While technology is a common focal example for the diffusion of innovation, other innovations can be described by the same diffusion process, though not all innovations will share the characteristics of interest here. For example, organizational governance practices such as compensation schemes are innovations that spread through inter-firm social and board interlock networks and are shaped by matching incentives (Davis & Greve, 1997). Where governance practices are shaped by matching incentives and offer different performance benefits, they can be described with the theoretical framework studied here. However, research on the diffusion of innovation also addresses behaviors that are outside the present theoretical scope, since innovation adoption doesn’t always involve matching incentives, as for example the many different types of computer displays—people don’t care about matching and conventions don’t emerge. Similarly, not all conventions are germane to the present analysis since value is not always a concern. For example, people don’t care whether they drive on the left or the right side of the road as long as they all agree.

It is worth briefly discussing other related ideas that share similar characteristics to technological conventions and yet represent distinct social phenomena. Research on “dominant design” addresses a wide range of mechanisms which are distinct from the process studied here but can lead to the widespread adoption of a single technology (Suarez, 2004). For example the term “standard” often invokes some intentional, centralized process such as a committee, not the emergent byproduct of individual optimizing as studied here. The diffusion of innovations is also distinct from the diffusion of beliefs and ideas, although this related research does provide conceptual support to the idea that network structure can shape the optimality of group behavior. For example, the speed of information exchange can impact research and development, where network centralization can reduce the breadth of exploration (Lazer & Friedman, 2007). Similarly, beliefs about things such as economic forecasts are shaped by network structure, and centralization reduces accuracy (Becker, Brackbill, & Centola, 2017). However, these processes represent fundamentally different cognitive and social behaviors. For example, information exchange and belief formation are shaped by effects such as anchoring and bayesian updating, which represent distinct mechanisms from the matching incentives shaping innovation adoption.

**Networks, Incentives, and Diffusion**

The added value that adopters get by matching peers is known as a "coordination externality," which is a subclass of a more general incentive known as a "network externality" (Arthur, 1989; Katz & Shapiro, 1985). Notably, the term "network externality" is somewhat a misnomer with respect to network theory in the contemporary sense because network externalities in general are often modeled using only total market adoption without consideration for the actual structure of organizational networks (Lee et al., 2016). Importantly, however, coordination externalities represent the added value that potential adopters get specifically from matching with the firms or people in their own network, rather than the market as a whole, and thus modeling network structure is important for explaining the diffusion of innovation (Ellison, 1993; Lee et al., 2016). For example, railroad companies want to ensure that their rails are physically compatible with partner firms, but this interest is limited to geographically proximate railways (Puffert, 2002). Programming language choice is shaped by direct peer interaction at conferences (Fang et al., 2021), and communication technology in general is motivated by the peers with whom one wishes to communicate. Thus modeling networks is important.

Coordination externalities can derive from multiple mechanisms, since any generic matching incentive can be formally modeled as a coordination externality. In the context of technology, one of the most straightforward coordination incentives is the literal need to coordinate as in the case of technologies where compatibility matters, exemplified by the development of technological conventions in railroads (Puffert, 2002). More generally, matching incentives can be used to represent generic copying or "mimetic" behavior, where firms copy each other for a range of reasons, for example as a method of uncertainty reduction (DiMaggio & Powell, 1983). One reason mimetic incentives can shape innovation diffusion is that popularity can be a heuristic for payoff: when one firm observes another's choice to adopt some technology, the observing firm infers that the adopting firm has received information indicating that the focal technology is the best choice (Greve & Seidel, 2015).

One key scope consideration for research on diffusion is the type of network through which the innovations spread. While innovations often spread through firm-to-firm networks as in the examples of railroad technology or organizational structure, the diffusion of innovation and the emergence of conventions is not limited only to inter-firm networks, and innovations can also spread through individual person-to-person networks. As noted above, software development language choice spreads through peer-to-peer networks at professional conferences (Fang et al., 2021). Conventions can also emerge between people within organizations. For example, Wikipedia and OpenStreetMap represent decentralized organizations of volunteers who must maintain a high level of coordination—editors must all produce a similar product. While such practices are codified, this documentation only formalizes conventions that emerge endogenously (Hall, Thebault-Spieker, Sen, Hecht, & Terveen, 2018; Rijshouwer, Uitermark, & de Koster, 2021). The present paper studies a generic model of adoption in networks which can represent diffusion either within or between organizations.

The present paper builds on previous research examining the effect of network structure on the diffusion of innovation, but this research focused on topics like speed and not the optimality of competitive dynamics. For example, theoretical modeling very similar to the present work has identified how network structure impacts the speed of diffusion when a superior innovation is introduced against an established convention (Montanari & Saberi, 2010). Similar models have been used to examine how network structure impacts the strength of lock-in effects once established (Lee et al., 2016). A great deal of research has focused on the role of central individuals, both establishing the relationship between network centrality and the ability to spread innovations further and faster (Banerjee et al., 2013; Coleman et al., 1957) as well as attempting to determine which metric of centrality best predicts influence in the adoption process (Banerjee et al., 2013; Guilbeault & Centola, 2021). However, this research focuses on the process of one new thing spreading, and not the path-dependent emergence of a convention from multiple competitors. The present paper examples how network structure impacts competition.

**Modeling the Diffusion of Innovation**

Empirically studying path dependence is inherently constrained by the inability to know whether history might actually have turned out differently, and formal computational modeling provides a strategy for studying path dependent processes (Vergne & Durand, 2010) such as innovation diffusion (Puffert, 2002). For any given empirical case, it is endlessly debatable whether something could have turned out differently, i.e. whether a technology's dominance was inevitable or historic chance. For example, QWERTY keyboards are often upheld as an example of a technology shaped by path dependence (Arthur, 1989; David, 1985) even as others argue that it was the best and only possible outcome (Kay, 2013). Where empirical identification is limited, formal models provide a form of "virtual experiment" (Macy & Willer, 2002) that can be used to develop and test hypotheses deriving from simple approximating premises (Epstein, 1999). In the present context, formal game theoretic coordination models (Young, 1993) can be used to test how network structure impacts the likelihood of optimal conventions.

A simple but powerful model of group-level innovation diffusion can be developed by starting with just a few basic approximating premises about individual behavior. A formal definition is provided below, which can be summarized as follows: people adopt the boundedly rational best response to their observation of peer choices. People choose the behavior or technology with the highest expected value, where expected value combines the number of peers using a given strategy (the coordination incentive) with the inherent payoff or value of that strategy. This model represents a context where there is no value at all to a strategy employed by no peers, as in a telephone service with nobody to talk to. This model necessarily represents an approximation of empirical behavior, since no real-world instance of any phenomena matches any model exactly. However, this limitation is also a strength: this model provides a method for assessing the effect of coordination incentives, holding all other effects constant. This model explains how arbitrarily large groups can spontaneously converge on shared convention and is thus a useful model for studying the diffusion of innovation and the adoption of suboptimal technologies (Ellison, 1993; Young, 1993).

An important context for the present research is that the model studied here is a general model of competitive diffusion that is formally identical to a basic contagion model, i.e. the case where one innovation is introduced into a population with an existing convention. Thus, the spreading of a single new innovation can be theorized as a special case of a more general behavior that leads to the emergence of conventions. When the relative payoff of a new innovation is very high compared to the established convention, then the model reduces to that of a "simple" contagion i.e. a standard epidemiological model: a person will adopt as soon as they learn about the innovation from even one peer. Alternatively, the relative payoff may be close enough to the established convention that one requires several peers to adopt before it is worth switching. In such a scenario, this innovation diffusion model is formally identical to a "complex contagion" model in which people only adopt an innovation if some sufficiently large proportion of their peer group also adopts (Centola & Macy, 2007).

These related models don't directly explain conventions, but their validation by empirical research provides valuable context for the present paper. Although the study of social and technological conventions is limited by inherent empirical constraints as described above, prior research on the diffusion of innovation has provided empirical support for the general class of model studied here. For example, a field study on the diffusion of microfinance supported the theoretical prediction that central individuals can spread innovations further and faster in a network (Banerjee et al., 2013), a core principle of the present paper. A controlled experiment on the promotion of healthy difficult-to-spread healthy behaviors shows that network structure interacts with peer effects to shape adoption (Centola, 2010), another key feature of the present analysis. While these prior studies don't reflect the competition of multiple strategies, they nonetheless provide encouragement that the class of model studied here is empirically valid.

## METHODS (MODEL DEFINITION)

This paper studies a game theoretic coordination model (Lewis, 1969; Morris, 2000; Young, 1993). A coordination model can be described in the following way: a population of decision-making agents must frequently perform some behavior, e.g. writing computer scripts, where they must be compatible with others, e.g. to share scripts. They must choose a strategy, e.g. a coding language, by combining two pieces of information: the payoff of each strategy (i.e., the inherent value) and the number of network neighbors using that particular strategy. Agents choose the strategy that offers the highest expected payoff (the "best response" strategy) which combines the need for coordination (e..g compatibility or conformity) with the payoff of the strategy. Agents periodically update, changing their strategy if their peer behavior changes.

### Formal Definition

This paper examines an MxM pure coordination game (i.e., with *M* possible strategies) in which a strategy $S_i$ offers payoff $V_i$ upon successful coordination, and a payoff of 0 upon miscoordination (failure to coordinate). The expected payoff for strategy $S_i$ is equal to

*(proportion of neighbors using $S_i$) x (payoff of $S_i$)*

which follows the standard calculation for expected payoff. Number of neighbors can be equivalently substituted for proportion of neighbors.

This expected payoff function offers several possible interpretations in the context of coordination behavior. In a context where agents play a series of pairwise (two player) coordination games, then the proportion of observable peers employing a given strategy represents the probability of success on a given pairwise interaction. Under one interpretation, an agent interacts only with their network neighbors. For example, a firm choosing a communication protocol must attempt to match as many possible peers in their immediate business network. In another interpretation, agents may have some chance of interacting outside their network neighborhood, but use observations of their immediate network neighborhood as a heuristic to form beliefs about the population at large.

Following this model, each run of the simulation is generated as follows:

- N agents are embedded in an undirected, binary network
- Initially, each agent is randomly assigned a strategy from a set of M possible strategies.
- At each time step, an agent is randomly selected from the set of agents whose strategy is not already the best response to their social environment.
- The focal agent adopts the best response strategy (best response to peer behavior) with probability (1-$\varepsilon$). With probability $\varepsilon$, the agent randomly selects a peer and adopts their strategy.
- The model is run until all agents are employing a best-response strategy.

This model can reach an equilibrium where multiple strategies are in use throughout a population (Blume, 1995). However, such outcomes are rare, and the networks presented here reach a global convention in more than 99% of simulations.

**Model Parameters**

The primary behavior of interest here is the case in which a population must select an equilibrium from a large number of possible strategies. To study these dynamics, the main results presented here model the case where there are many possible strategies than people (M=N), with additional simulations shown in the Appendix producing qualitatively similar results. The Appendix provides additional details regarding other parameters such as the distribution of strategy payoff and noise parameters. For centralized networks, results in the main text present networks with an average of 20 edges per actor. All networks presented here have N=1000 actors, and results are comparable for larger and smaller networks.

**Network Structure**

A network is considered "centralized" when there is inequality in the distribution of connectivity, i.e. when some actors are better connected than others. In the most highly centralized networks, one or a small number of actors have a large number of connections, while most actors have only a few connections (Freeman, 1978). This analysis quantifies the level of centralization in a network with the Gini coefficient. In the case of networks, the Gini coefficient is measured for the degree distribution (Badham, 2013), where each actor's "degree" is defined as the number of connections they have to other actors (Easley & Kleinberg, 2010). To produce centralized networks, this analysis employs two network generating algorithms. In order to provide a robust analysis and generate highly centralized networks across a wide full range values for the Gini coefficient, this paper employs two distinct network generating mechanisms, detailed in the Appendix. In the results presented below, the two network types are represented separately in the figures.

Prior work has shown that regular lattice networks where actors are connected in a perfect grid do not allow for lock-in on suboptimal strategies, since even marginally superior strategies generate cascades which spread quickly through the entire network (Ellison, 1993; Blume, 1995). However, real-world networks are more complex, not geometrically regular (Watts & Strogatz, 1998) and do generate lock-in effects that can lead to suboptimal conventions(Ellison, 1993; Montanari & Saberi, 2010). Even minor disruptions to the geometric regularity can disrupt cascading effects (Centola & Macy, 2007). This paper therefore focuses its analysis on complex networks, not regular graphs.

## RESULTS: NETWORK STRUCTURE AND EQUILIBRIUM SELECTION

This analysis examines the effect of centralization on the probability of optimal coordination, showing that centralized networks are more likely to converge on suboptimal solutions. In order to test whether the effects of centralization can be explained as a result of a speed/optimality tradeoff, this analysis also tests the effect of network efficiency by varying the density of decentralized networks. The Appendix provides additional analysis showing the robustness of the main results against variation in modeling assumptions including a smaller set of possible strategies and variation in network assumptions.

**The Effect of Network Centralization**

Figure 1A shows the probability of optimal coordination in random networks with increasing centralization, holding constant network density (average degree). In decentralized networks, the population selects an optimal strategy in about 70% of simulations. As centralization is increased, however, the probability of optimal equilibrium selection decreases dramatically. In the most centralized networks, the optimal solution is selected in only about 15% of simulated trials. Points in black and grey points show two different network generating algorithms. Points average across 5,000 simulations, and 95% confidence intervals are drawn but too small to be visible on most points.

---------------------------

Insert Figure 1 About Here

---------------------------

One possible explanation for this reduced optimality is a speed/optimality tradeoff, since Figure 1 (right) shows that speed of convergence increases with centralization. However, further analysis suggests that this effect is due to the ability for central actors to cause low-payoff strategies to spread. One simple way to illustrate this effect is to consider whether central actors are more likely than others to introduce the strategy that goes on to become the emergent convention. If central actors themselves adopted the optimal strategy from the environment, and then spread that strategy, then central actors would be no more likely than anybody else to be the initial introducer of the strategy that becomes the convention. However, as shown in Figure 2A, central actors are more likely to introduce the winning strategy than would be expected by chance. This figure shows the likelihood ratio L(D) comparing the probability that an actor of degree D introduces the winning strategy in the simulation results, against the baseline probability of introducing the winning by sheer chance. Outcomes are measured across 100k simulations of a single centralized network, see Appendix for details. Thus, influence beats information access, and central actors tend to simply spread whatever solution they initially introduce.

---------------------------

Insert Figure 2 About Here

---------------------------

Another way to illustrate the effect of centralization is to examine the relative payoff of emergent conventions, conditional upon the optimal strategy becoming the convention. To motivate this analysis conceptually, consider the case where the optimal solution is only marginally better than the second best solution. In a decentralized network, every solution has an equal chance to spread, and the optimal solution will spread. In a centralized network, however, that optimal solution must compete against the influence of solutions introduced by central actors. As a result, optimal solutions in centralized networks will only spread if they hold a relatively larger advantage. This explanation is illustrated in Figure 2B, which shows the average payoff of the emergent convention as a function of network centralization, conditional upon being the optimal strategy (dashed line) or the suboptimal strategy (solid line). Overall, centralization reduces the average payoff due to reduced optimality (solid line). However, considering only those cases where the emergent convention was optimal (dashed line) centralization increases the payoff. This result illustrates that in order for the optimal strategy to spread widely, it must offer a relatively higher payoff to succeed in centralized networks. This supports the argument that the effect of centralization is not a speed-optimality tradeoff, but rather a competition between innovations that are optimal-but-peripheral and those that are central-but-suboptimal.

One additional way to illustrate the influence of central actors is to compare the average payoff of equilibrium solutions introduced by central actors with those introduced by peripheral actors. As noted above, solutions introduced by central actors are more likely to be adopted, regardless of merit. The complementary expectation is that the payoff for a solution introduced by a peripheral actor—conditional upon becoming the equilibrium solution—is expected to be higher than the payoff of an equilibrium solution introduced by a central actor. This relationship is confirmed in simulation (see Figure A1in the Appendix), such that successful strategies introduced by peripheral actors offer a greater payoff than strategies introduced by central actors. This result shows that in order for a peripheral actor to introduce the emergent convention, it must offer an even greater relative payoff in order to overcome the influence of central actors.

**Network Efficiency: Testing a Speed-Optimality Tradeoff**

While the analysis above casts some doubt on the speed-optimality tradeoff, it does not directly test whether it is possible for networks to be both fast and optimal. However, this tradeoff is not unavoidable, and in very dense networks, coordination is both fast and optimal. Network density counts the total number of connections as a fraction of total possible connections, and is proportional to the average number of connections per agent when the network size is held constant. Figure 3A shows that optimality increases with network density in decentralized networks, where the leftmost point in this figure is equivalent to the leftmost point in Figure 1A, showing networks with 20 connections per agent (10,000 simulations per point, 95% confidence intervals). In fully connected networks, groups nearly always select an optimal equilibrium. Increasing graph density not only increases the probability of optimal coordination, but also increases the speed of coordination. Figure 3B shows the average number of updates that each actor is required to make in order for a population to reach equilibrium, and is comparable to Figure 1B.

-----------------------------

Insert Figure 3 About Here

----------------------------

The explanation for the double advantage of network density (for both speed and optimality) can be illustrated by considering the trivial case of a fully connected network, in which every agent observes every other agent. At the outset, every actor has a unique solution. Because every actor has full knowledge of the strategy space, and no strategy yet has any advantage of popularity, every actor will simply adopt the best strategy. This argument also holds for the more general case where there are fewer strategies than people. In expectation, each strategy will be employed by the same number of people at the outset – thus, again, no strategy starts with a popularity advantage, and again the best strategy will be immediately adopted. Although the case of the fully connected network describes the trivial (and perhaps unlikely) scenario in which every agent has full knowledge of the network and thus the solution space, it serves as an illustration for the more general advantage of network density.

## DISCUSSION

This paper contributes to research on the diffusion of innovation by showing how a structural analysis of emergent conventions can be important to understanding market behavior, and that one feature determining the optimality of conventions is network centralization. If explanation stops with the conclusion that conventions are simply the result of “historical idiosyncrasy,” i.e. that emergent conventions are fundamentally inefficient, then interventions will only be aimed at innovations themselves by seeking to identify and spread the optimal choice before lock-in occurs. In contrast, this paper shows that inefficiency is not a necessary feature of emergent conventions, and that some conditions allow the best innovations to surface naturally through the invisible hand of the market. Thus instead of standards setting committees or other centralized interventions, future research may be able to identify structural interventions which seek to cultivate collective intelligence in the system as a whole. Such structural interventions could have the advantage not only of harnessing people’s natural behavior to identify optimal innovations, but also of generating lasting benefits versus single innovation interventions.

Admittedly, the present research largely identifies a problem—a reason why markets are inefficient—and falls short on solutions regarding how to actually shape networks, and identifying structural effects is easier than developing structural interventions. But following the effects theorized here, future research may be able to identify methods to mitigate the effect of central actors. This could occur for example by increasing the influence of peripheral participants, thus allowing for more equitable competition to surface the best option; or by seeking to encourage the creation of ties between peripheral participants, directly reducing centralization. Rather than targeting networks themselves, it is also possible that simply increasing information availability may mitigate the effects observed here. Because suboptimal coordination is made worse by the dominance of central individuals, people may prefer to take a short term cost of mis-coordination to adopt a superior alternative, if they know it exists, and if they know that others in the global network are likely to do the same.

As with all formal models, this theoretical analysis necessarily presents an approximation of "real" human behavior, and innovation adoption can rarely be fully characterized by a simple coordination game. However, this paper reveals the effect of network centralization to be an important source of market inefficiency that is nearly always likely to be impacting outcomes even if other co-occurring processes moderate the effect, since networks empirically tend to be centralized (Barabási & Albert, 1999). Moreover, this model has been successful at predicting results of laboratory experiments (Centola & Baronchelli, 2015; Centola, Becker, Brackbill, & Baronchelli, 2018) and the formally identical contagion model has been effective at describing field data (Banerjee et al., 2013) suggesting ecological validity for this general class of models.

**Limitations and Future Research**

One notable simplification relevant to the present study is the implication that coordination incentives, modeled here, are the only externality shaping behavior. However, even in the simple railroad example, there are some additional network externalities e.g. investment in equipment production facilities. Another limitation is the tendency to converge on a single population-wide choice, a phenomenon that is rarer empirically. Even railroad technology continues in several different forms, shaped in part by geographical lines, while other technologies such as smartphone choice segment markets in other ways. But the present model nonetheless offers theoretical insight. For example, cases where the population converges on multiple conventions, subcommunities (e.g., geographical region) may be modeled as following a separate coordination process, but future research should more carefully examine the effect of multiple subcommunities or scenarios where multiple externalities interact.

While models are always inherently limited, and most empirical cases will have many more incentives and processes shaping adoption, we can expect the effect of network centralization to operate as a background force alongside those effects. With respect to the possibility of multiple incentives shaping adoption behavior beyond just coordination externalities, we can expect other network externalities to simply enhance the effects observed here, since they operate in the same direction: both coordination and network externalities generate cumulative advantage for early success. All else being constant, we should expect centralized networks to be suboptimal compared to decentralized networks, and the key question for future research is whether there are any processes that could mitigate this effect. For example, future research examining interactions with other network externalities may reveal interventions that can mitigate or eliminate the deleterious effect of centralization. Such interactions, if identified, could point the way toward potential interventions or to market improve efficiency.

Although the coordination model presented here captures a highly stylized description of human behavior, the effects described here follow generally from the empirically supported expectations that central actors are influential (Banerjee et al., 2013; Katz & Lazarsfeld, 1955; Rogers, 1962) and that decisions are shaped by conformity pressure (Asch, 1951; Centola & Macy, 2007) or mimetic incentives (DiMaggio & Powell, 1983; Greve, 1996). Any time an influencer contributes to the popularity of a behavior or technology and people want to match their peers, there is an increased chance that the ideas introduced by central actors will become popular at the expense of superior strategies introduced by peripheral actors.

## Conclusion

The use of a generic theoretical model of innovation diffusion allows the formal results of this paper to inform a wide range of contexts, since “innovation” may be interpreted as many things including technology as hardware, practice, and behavior—a new dance move is an innovation, as is a new governance system, or a new communication protocol. Because all of these types of innovation are studied with the same network models, then any general principles arising from these models—e.g. the effect of network structure when multiple innovations compete—will apply equally wherever the model’s assumptions are met. This research thus contributes to theory focused on the adoption of technology as hardware but also shows how network structure can shape the emergence of social and technological conventions more generally.

This paper illustrates how the structure of a group as a whole impacts the optimality of group-level outcomes, and thus shows how research on the diffusion of innovation can contribute to the growing body of theory on collective intelligence (Becker, Guilbeault, & Smith, 2021; Bernstein, Shore, & Jang, 2022; Lazer & Friedman, 2007; Woolley, Chow, Mayo, Riedl, & Chang, 2022). As the emergent selection of one outcome from multiple possible choices, the emergence of conventions can be interpreted as a form of emergent collective decision-making. The formal model employed here represents a stochastic markov chain, where the emergence of conventions is a form of game theoretic equilibrium selection (Kandori et al., 1993; Young, 1993). Thus an emergent convention represents a collective “selecting” an outcome, albeit through the autonomous behavior of individual actors and not any intentional or explicit decision-making process at the group level. The effects described here thus show how the emergence of conventions in networks is a form of collective intelligence, a set of processes where groups display macro-level behavior that cannot be explained in terms of individual psychology but can only be explained by features of the population as a whole (Bernstein et al., 2022; Woolley, Chabris, Pentland, Hashmi, & Malone, 2010; Woolley et al., 2022).

The finding that centralization reduces optimality is consistent with broader research on collective intelligence. For groups completing estimation tasks such as forecasting, network centralization undermines accuracy (Becker et al., 2017; Golub & Jackson, 2010) much like centralization in the present work reduces the optimality of emergent conventions. While estimation tasks are a fundamentally different social process than innovation adoption, central actors face a similar informational situation in both scenarios. For estimation judgements, just as in choosing a technology, a central actor is in a position to integrate the information provided by a group as a whole. However, the peripheral actors are simultaneously influenced by the central actor, and the network is thus drawn toward the belief of central actors (Becker et al., 2017), much like with innovation adoption. The present research thus highlights a potentially general effect in collective processes whereby the deleterious effect of centralization is the influence of the central actor overshadows any potential beneficial effects of their position.

**Figure 1**

**Centralization Decreases Optimality (Panel A) and Increases Speed (Panel B)**

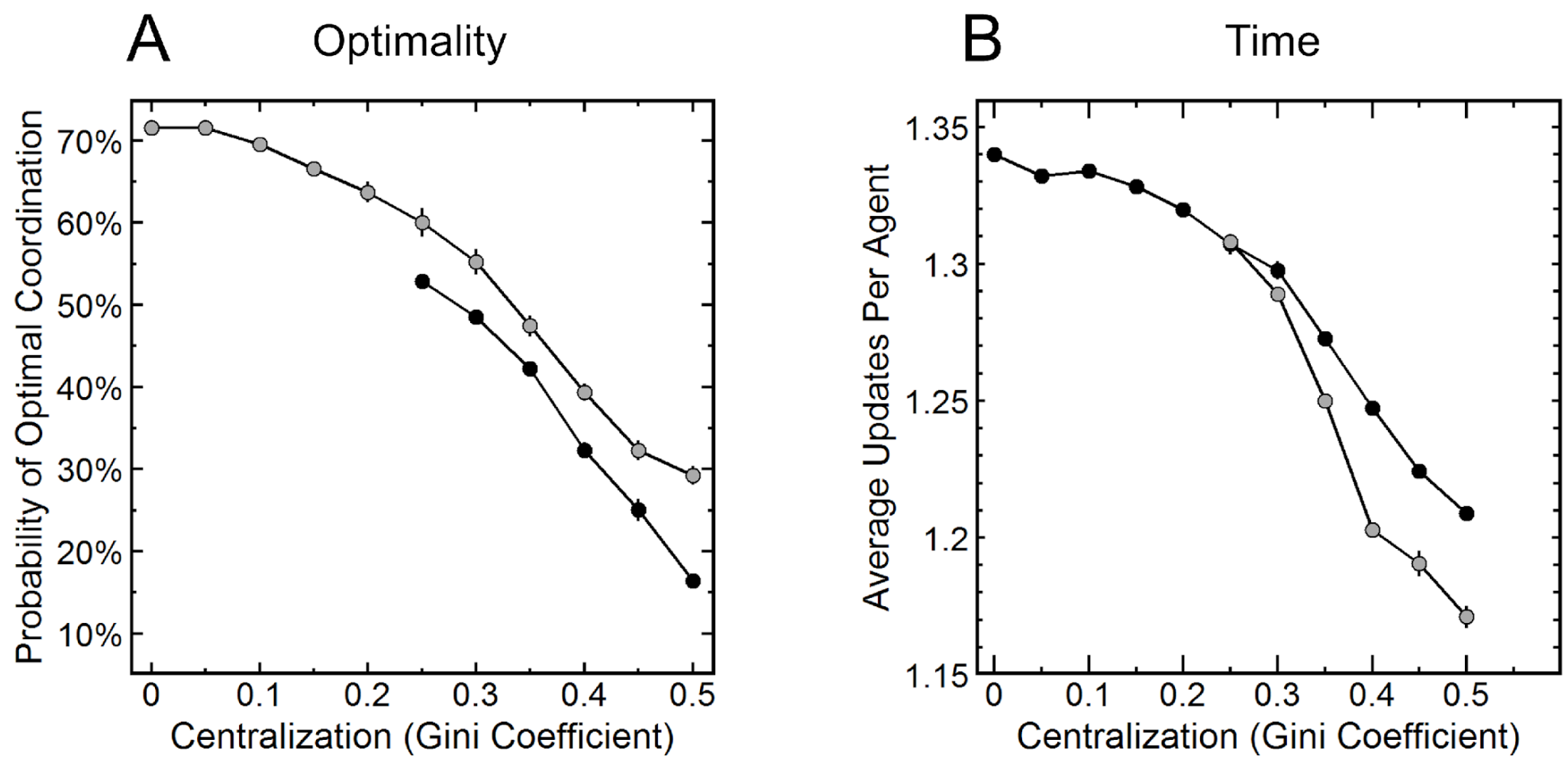

**Figure 2**

Centrality Gives Individual Strategies an Advantage (Panel A) but Centralization Increases the Difficulty of Optimal Coordination (Panel B)

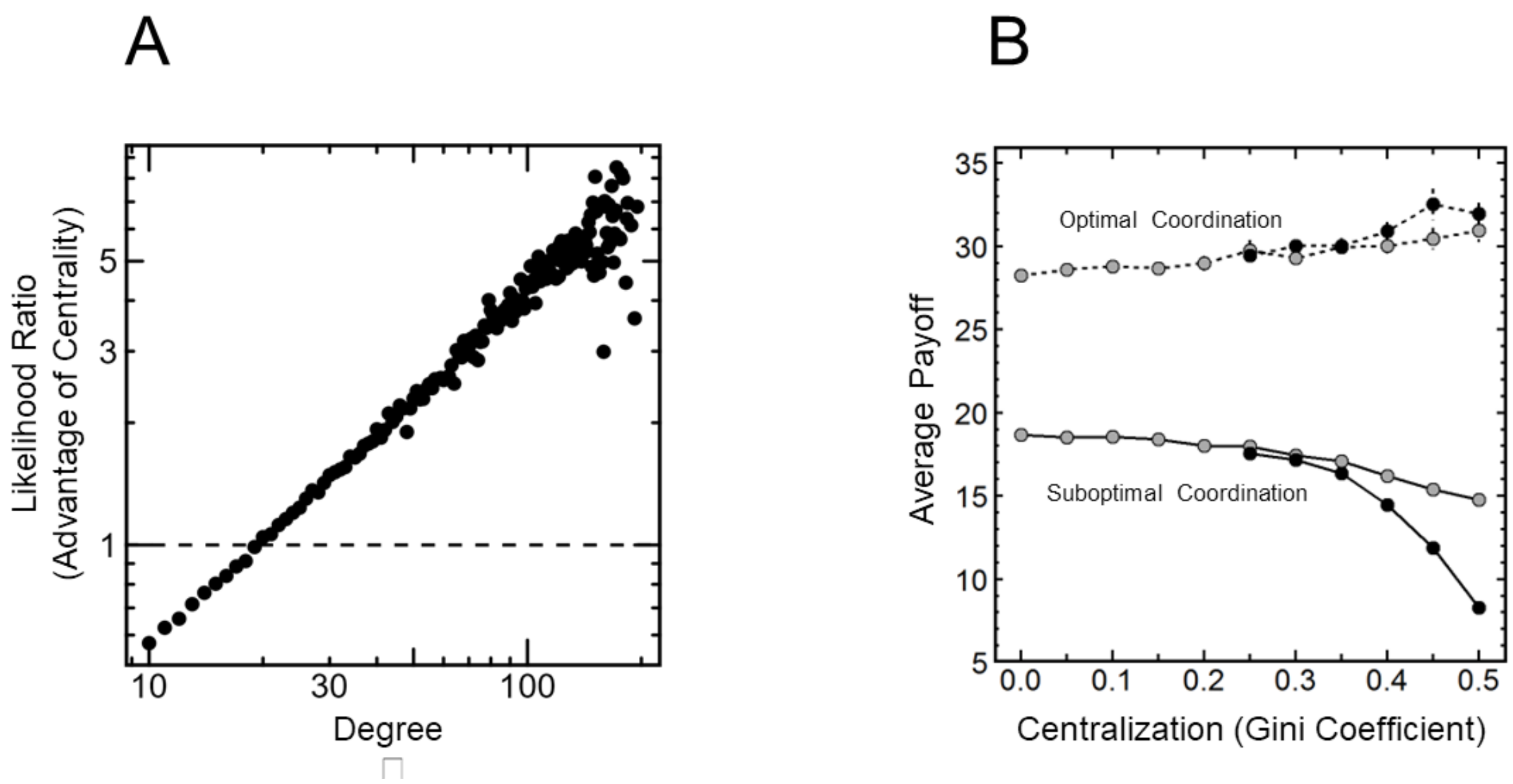

**Figure 3**

**Dense Networks are Both Fast and Optimal**

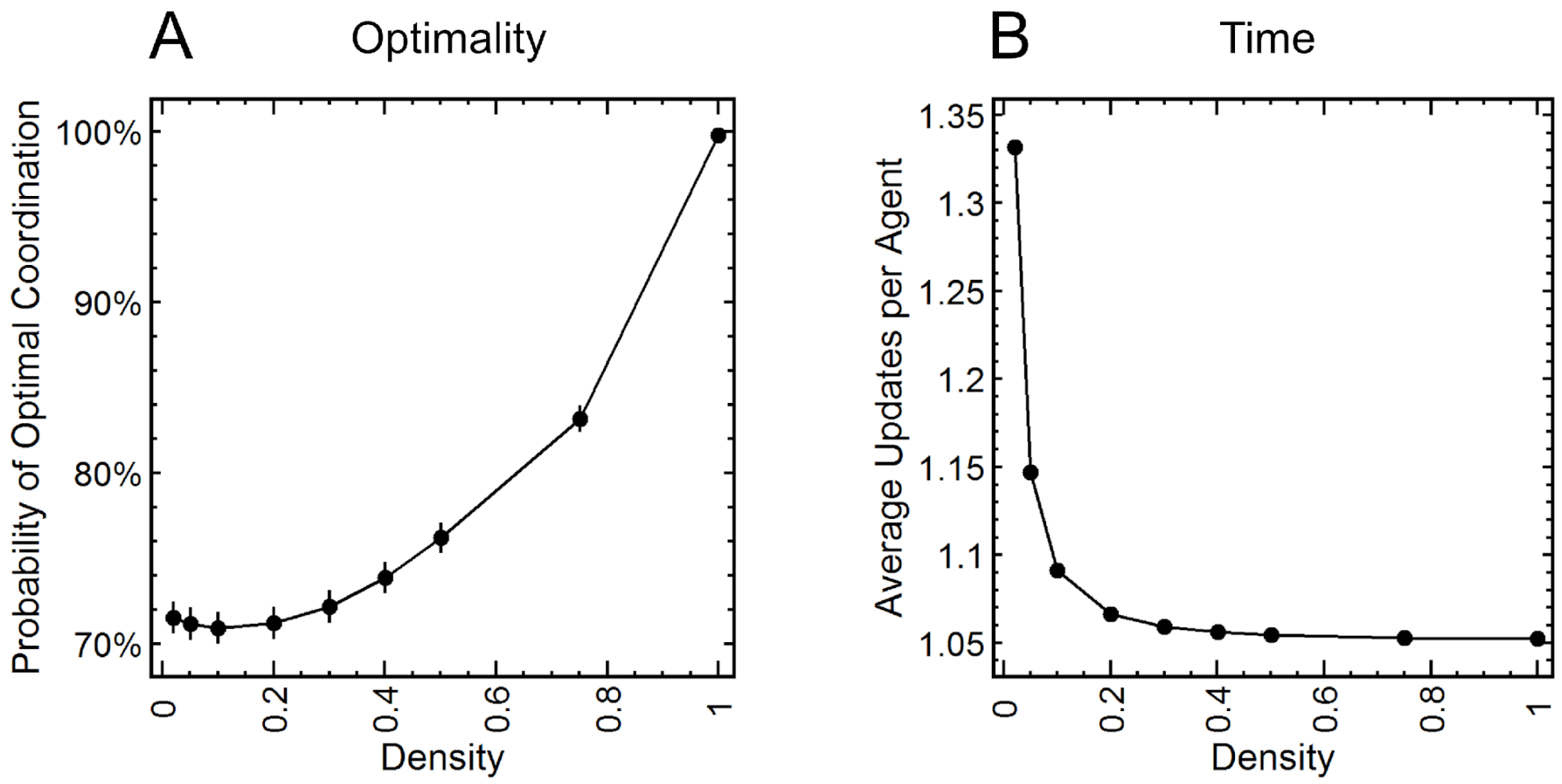

# APPENDIX

## Model Parameters

The payoffs for each of the M strategies are independently, identically distributed according to a fixed distribution. In the figures presented here, the payoff for a given strategy is drawn from a log-normal distribution ($\mu=0$, $\sigma=1$). The log-normal distribution is chosen due to the fact that innovations often follow a long-tailed payoff distribution, with many mediocre solutions and few high quality solutions (Kauffman, 1993). Qualitatively identical results obtain when strategies follow other distributions.

Results present outcomes with $\varepsilon=0.1$, meaning that 1 in 10 decisions is random, and 9 in 10 decisions follow the best response strategy. However, the effect of centralization on coordination does not depend on noise, and results are qualitatively identical when $\varepsilon=0$, such that agents follow a perfect best response strategy. For the sake of generality and consistency with previous research, this text presents results with noise.

## Network Generators

This analysis employs the popular “preferential attachment” mechanism developed by Barabasi and Albert (1999) to explain the emergence of scaling in empirical networks. When the parameters are appropriately tuned, this algorithm can generate networks ranging from moderately centralized random networks (Gini=0.25) to highly centralized “hub spoke” networks (Gini=0.5), in which a single core group of actors is connected to every other actor, while peripheral actors are connected only to the core actors. However, this algorithm cannot be tuned to generate fully decentralized networks (Gini=0). Therefore, in order to study networks that vary continuously between fully decentralized and highly centralized networks, this analysis employs a second algorithm, which generates networks based on an arbitrary degree distribution (henceforth "degree sequence" networks).

Preferential attachment networks are generated according to the algorithm developed by Barabasi and Albert (1999). To vary the range of centralization produced by this generator, I vary the strength of preferential attachment. As new actors are connected, they are connected to existing actors with probability

$$P \sim ck^{a} \qquad [1]$$

where $k$ is an actor's degree, $a$ controls the strength of preferential attachment and $c$ is a normalizing constant. This model reduces to the original Barabasi-Albert algorithm when a=1.

The preferential attachment generator produces networks with a Gini coefficient in the range of 0.25 to 0.5. To increase the range available for study, and to allow for the study of a continuous variation from decentralized (Gini=0) to centralized networks, I also generate networks using arbitrary degree distributions. To accomplish this, I first produce arbitrarily centralized degree sequences (random number distributions) drawing from an approximately power-law distribution with a range of parameters, producing degree distributions with Gini coefficients ranging from Gini=0 to Gini=0.5. I then use these sequences to produce networks with the Viger and Latapy (Viger & Latapy, 2005) method. When Gini=0, every actor has the same number of neighbors. The parameters are selected through a trial-and-error method to ensure a constant average degree (i.e., to vary centralization while holding graph density constant).

**Centrality Likelihood Ratio**

Because every actor has an equal probability of introducing the optimal solution, then the equilibrium solution should (optimally) be just as likely to come from a peripheral actor as a central actor, after controlling for the expected number of each type of actor. However, central actors are more likely to introduce the winning solution than would be expected by their prevalence in the population.

This effect is measured as follows. For any given simulation, let D be the degree (number of connections) for the "innovator," i.e. the agent that introduced the eventual equilibrium solution. If the equilibrium solution were determined only by payoff, or if equilibrium solutions were selected randomly, then the distribution of D would be equal to the degree distribution of the network itself. E.g., if only 1% of actors have degree 10, then only 1% of the equilibrium solutions should come from actors with degree 10. Since degree is a discrete variable (can only take integer values), the innovator degree distribution can be compared with the network degree distribution as the likelihood ratio

$$L(D) = \frac{proportion\ of\ innovators\ with\ degree\ D}{proportion\ of\ all\ actors\ with\ degree\ D}$$

where innovator refers to an actor that introduced the equilibrium solution. If the likelihood ratio L(D) is greater than 1, that means that actors with degree D are more likely than expected to introduce the equilibrium solution. If the ratio L(D) is less than 1, then actors with degree D are less likely than expected to introduce the equilibrium solution.

Figure 2A in the main text shows L(D) for 200,000 simulations on centralized networks generated with a preferential attachment algorithm and an average of 20 connections per actor.

These results do not depend on the particular shape of the network degree distribution, and similar results emerge for Erdos-Renyi random graphs, which have a Poisson degree distribution.

**Figure A1**

Figure A1 provides supporting information regarding the effect of centralization and the hypothesis that centralization reduces optimality by increasing speed. The payoff for a solution introduced by a peripheral actor—conditional upon becoming the equilibrium solution—is expected to be higher than the payoff of an equilibrium solution introduced by a central actor. Figure A1 indicates the payoff of conventions as a function of the centrality of the actor who introduced the strategy. This figure indicates an “underdog” effect, such that successful strategies introduced by peripheral actors offer a greater payoff than strategies introduced by central actors.

**Figures A2 to A6 (Robustness Analyses)**

The results in the main text demonstrate the properties of only one point in the parameter space with regard to statistical noise, payoff distribution, population size, and number of solutions. In particular, the simulations have assumed a large level of noise, an infinitely large strategy space (so that agents begin with individually unique solutions) and strategy payoffs drawn from a log-normal distribution. Effects are qualitatively unchanged by the absence of noise, variation in the payoff distribution for strategies, and changes in population size.

One factor that does moderate the effect of centralization is the size of the solution space. As the number of solutions is reduced, the effect of centralization becomes weaker, though it still has a substantial impact on the probability of optimal coordination. Even with only 10 unique solutions, the most centralized networks still select an optimal equilibrium less than 80% of the time, as compared with near-optimality in decentralized networks (see Figure A2). However, once the number of solutions reaches its limit (binary coordination, with 2 solutions) the effect of centralization decreases substantially. The decreased impact of centralization can be explained in terms of the initial conditions of the model. At time t=0, every begins with a randomly assigned strategy. When each strategy is unique, the strategies adopted by central actors have a structural advantage. However, with only two strategies, the structural advantage of each strategy will, in expectation, be approximately the same: each strategy is likely to be initially adopted by the same number of central individuals.

Although the effect of centralization is small in binary coordination, the most centralized networks nonetheless show a slightly decreased probability of optimal coordination as compared with decentralized networks. In the most centralized preferential-attachment networks, simulations converge on the optimal solution approximately 96% of the time, as compared with 99% optimality in the least centralized networks. The robustness of this effect shows that even in relatively straightforward two-choice decisions, the structural advantage given to solutions adopted by central actors always has some impact, if relatively small, on collective decisions.

Figures A3 to A6 replicate the figures from the main text examining the effect of centralization, but with half and double the average degree (K=10, K=40). These figures show qualitatively identical results.

**Figure A1**

**Winning Solutions Introduced by Central Actors Have Lower Payoff**

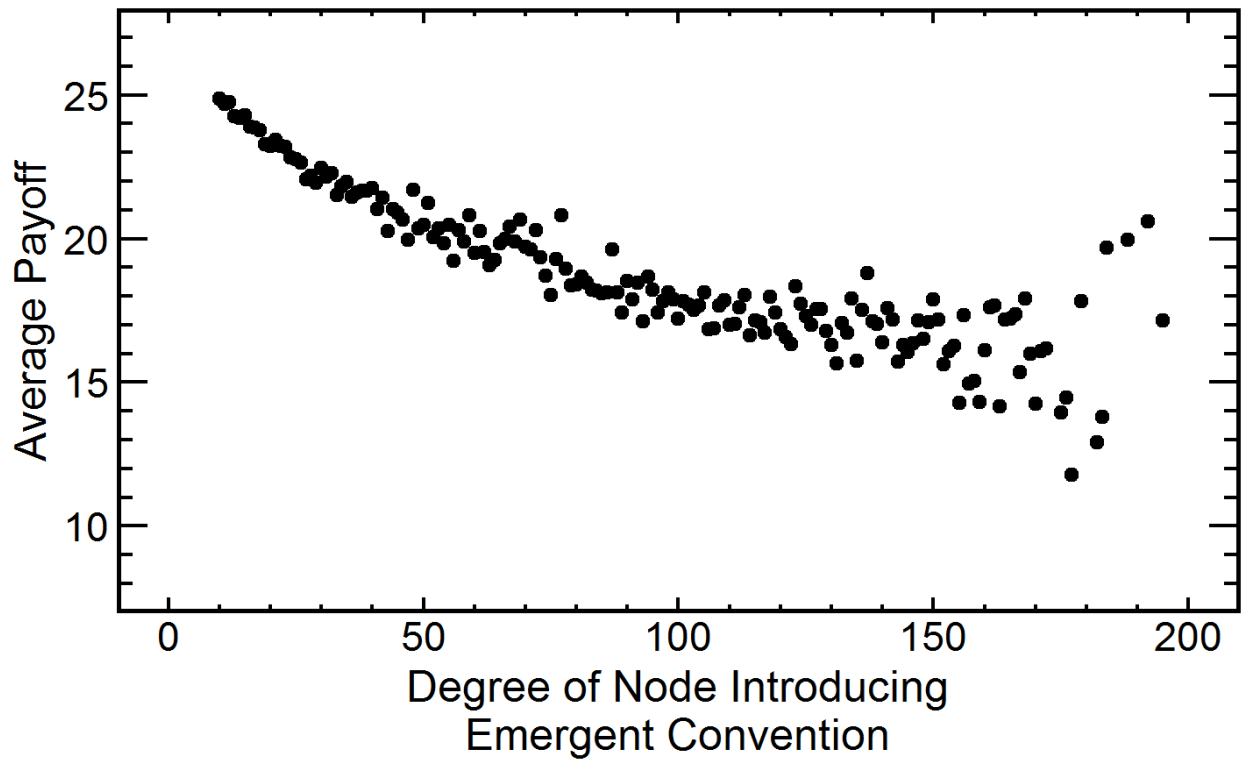

**Figure A2.**

**The effect of centralization is robust to the size of the strategy space**

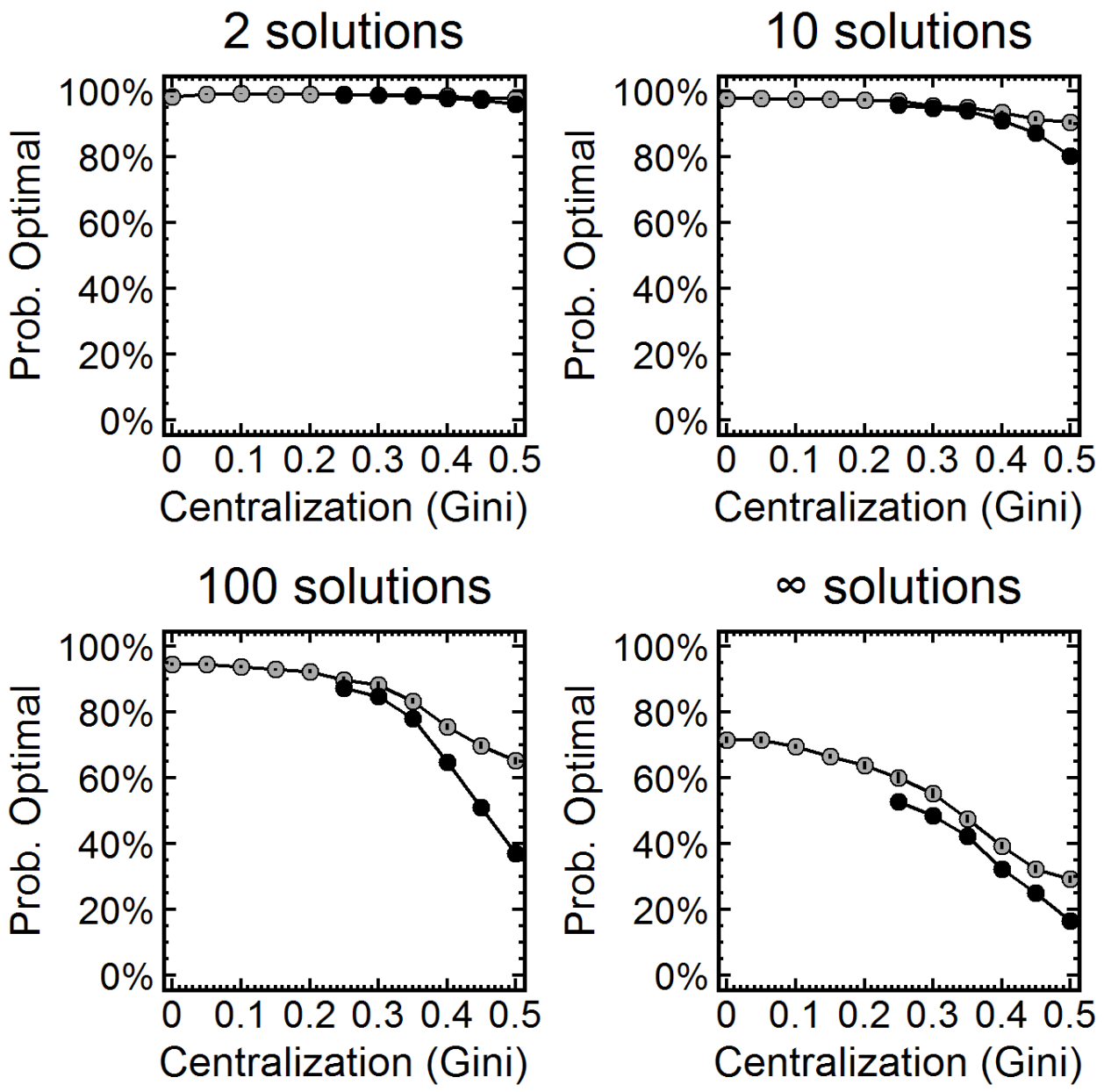

**Figure A3.**

**Replication of Figure 1 with varying average degree (K).**

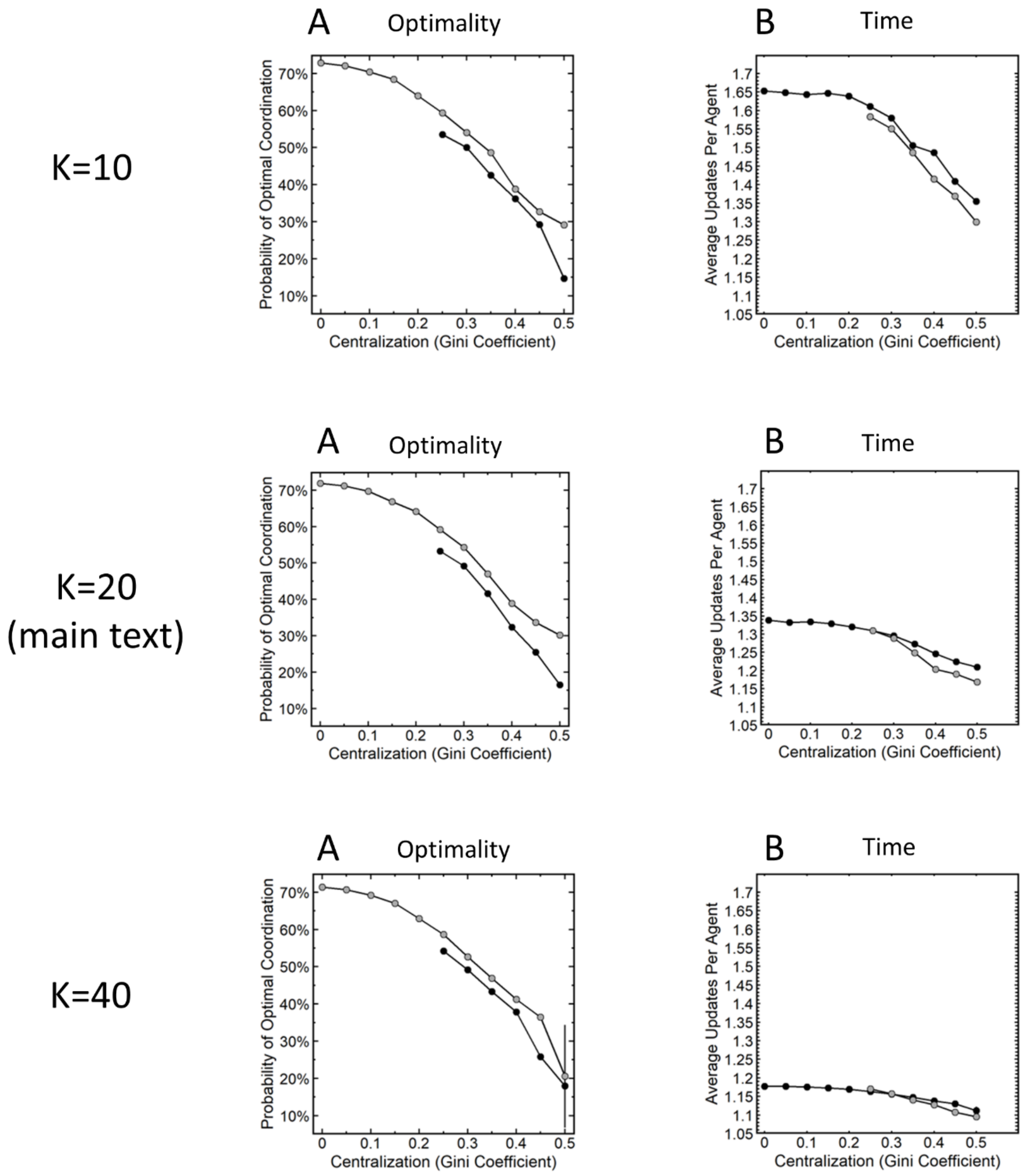

**Figure A4.**

**Replication of Figure 2A with varying average degree (K).**

K=10

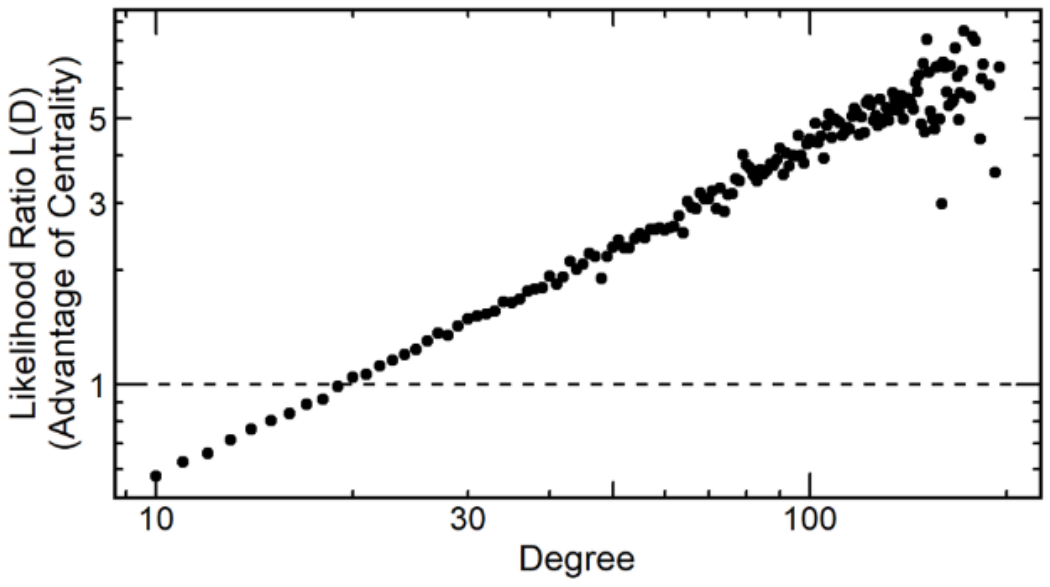

K=20
(main text)

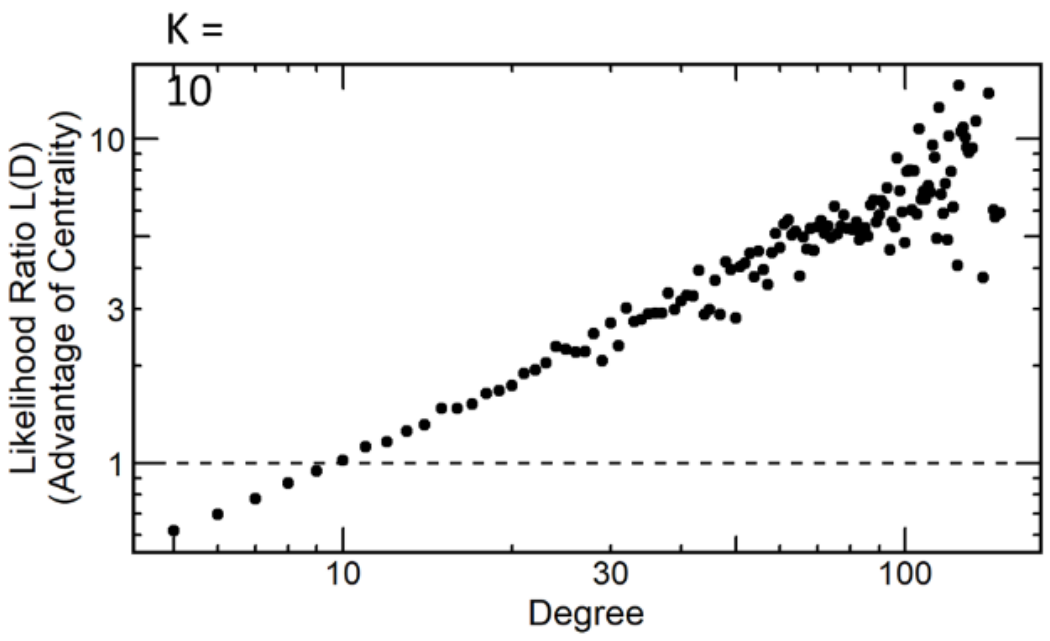

K=40

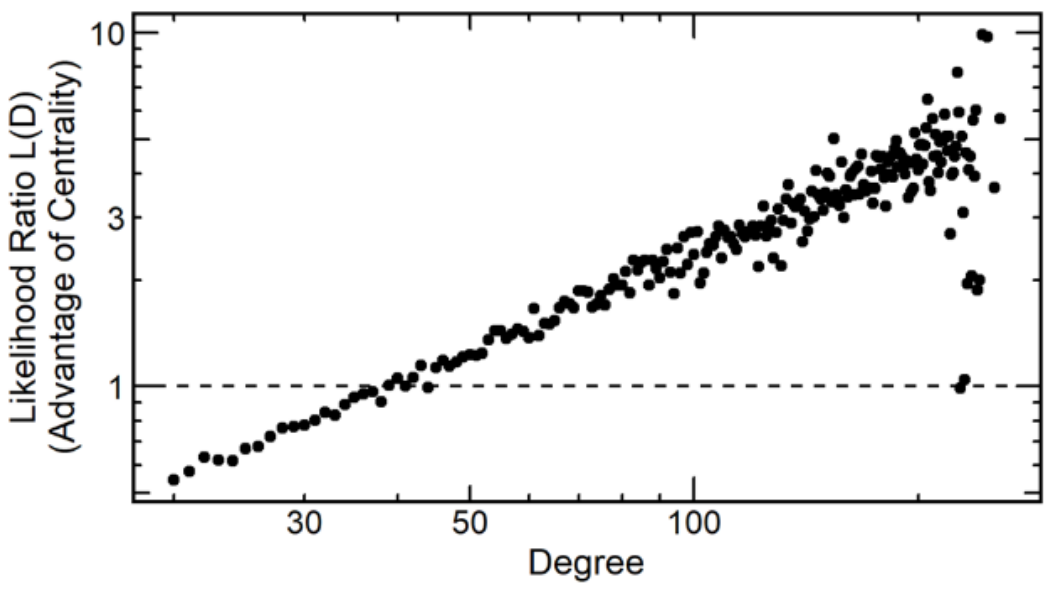

**Figure A5.**

**Replication of Figure 2B with varying average degree (K).**

K=10

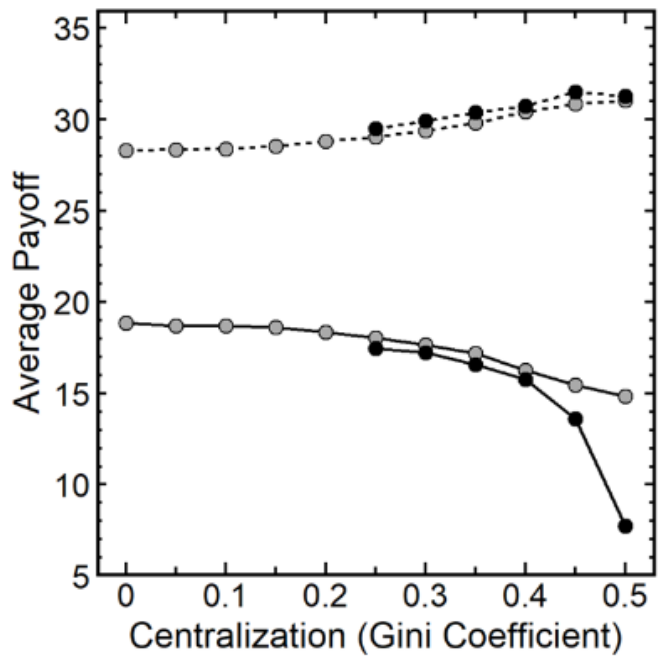

K=20
(main text)

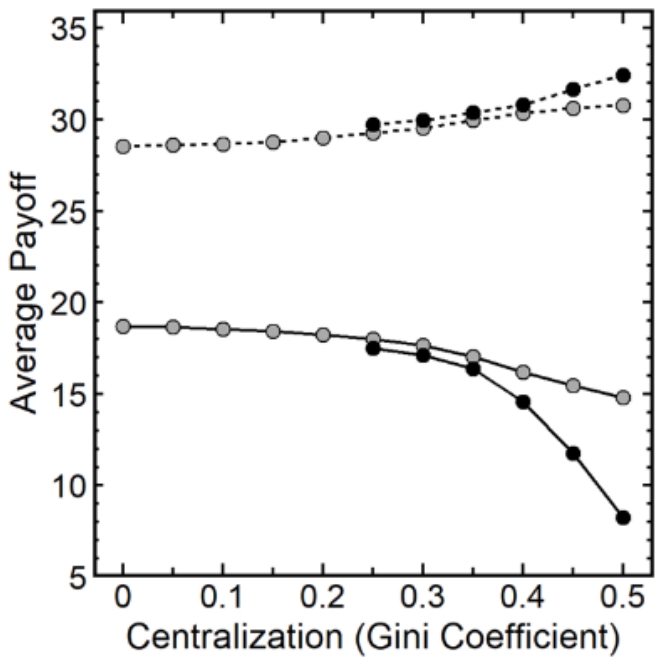

K=40

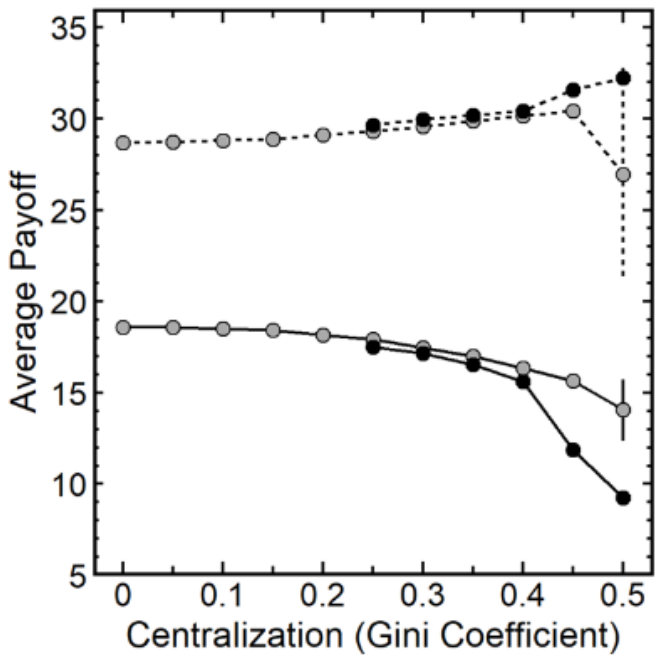

**Figure A6.**

**Replication of Figure A1 with varying average degree (K).**

K=10

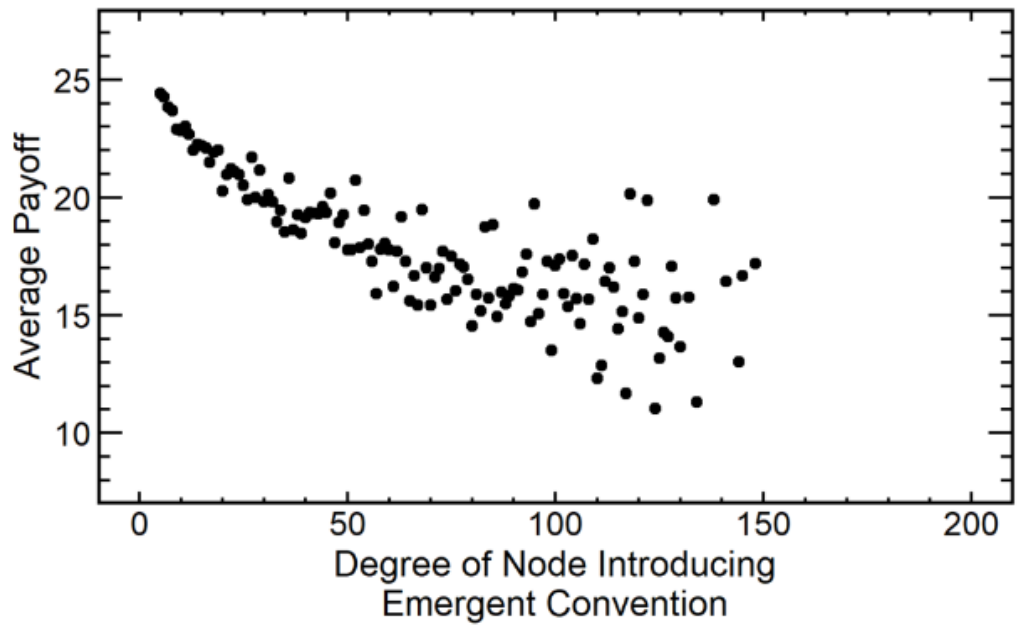

K=20
(main text)

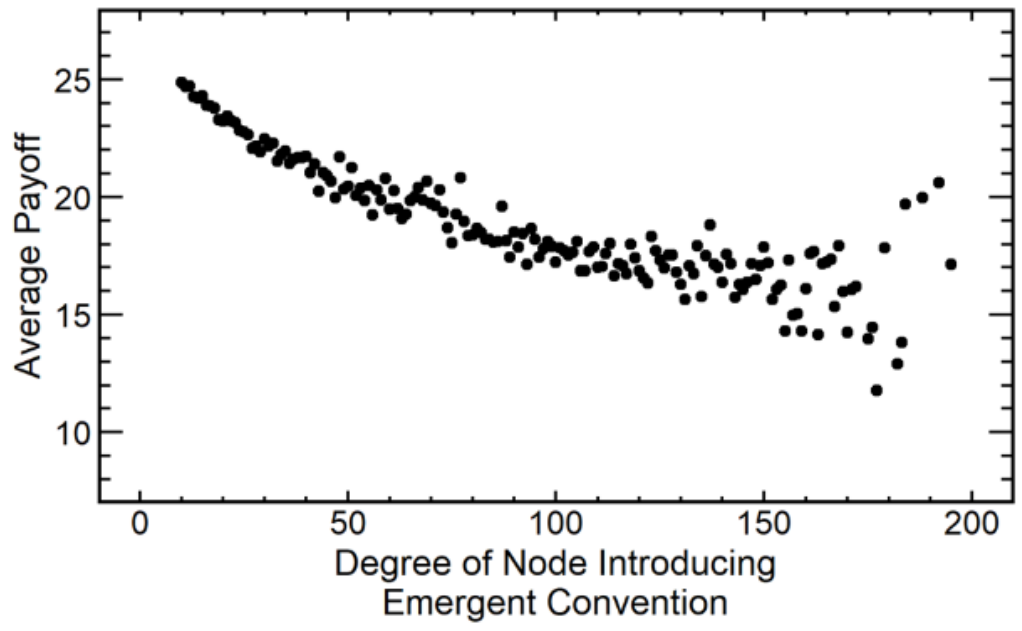

K=40

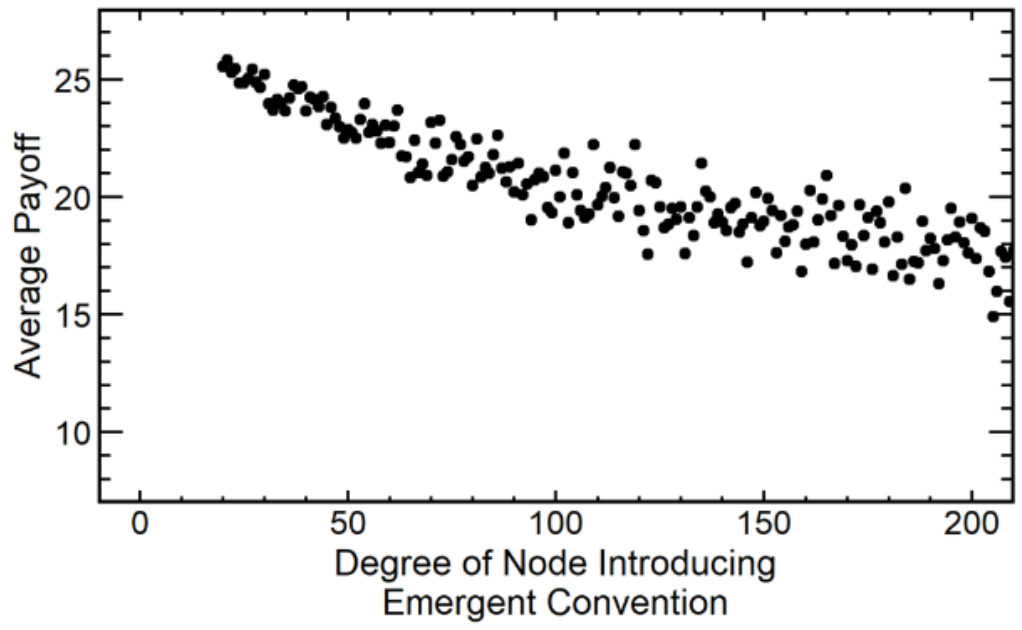